\documentclass[aps]{revtex4}

\usepackage{epsfig}
\usepackage{epstopdf}
\usepackage{color}

\usepackage{graphicx}
\usepackage{longtable}
\usepackage{CJK}

\begin{document}

\title{Quasi-exactly solvable cases of the $N$-dimensional \\ symmetric
quartic anharmonic oscillator}
\author{Feng Pan}\thanks{e-mail: daipan@dlut.edu.cn}
\affiliation{Department of Physics, Liaoning Normal University,
Dalian 116029, China}\affiliation{Department of Physics and
Astronomy, Louisiana State University, Baton Rouge, LA 70803-4001,
USA}

\author{Ming-Xia Xie}
\affiliation{Department of Physics, Liaoning Normal University,
Dalian 116029, China}

\author{ Chang-Liang Shi}
\affiliation{Department of Physics, Liaoning Normal University,
Dalian 116029, China}

\author{Yi-Bin Liu}
\affiliation{Department of Physics, Liaoning Normal University,
Dalian 116029, China}

\author{J. P. Draayer}
\affiliation{Department of Physics and Astronomy, Louisiana State
University, Baton Rouge, LA 70803-4001, USA}

\begin{abstract}

The $O(N)$ invariant quartic anharmonic oscillator is shown to be exactly
solvable if the interaction parameter satisfies special conditions.
The problem is directly related to that of a quantum double well
anharmonic oscillator in an external field. A finite dimensional
matrix equation for the problem is constructed explicitly, along with
analytical expressions for some excited states in the system.
The corresponding Niven equations for determining the polynomial
solutions for the problem are given.

\vskip 0.5cm
\noindent {\bf Keywords}: Quartic anharmonic oscillator;
quasi-exact solutions; Niven equation.
\vskip 0.5cm
\noindent {\bf PACS numbers}:~{03.65.Ge, 02.30.Mv, 02.20.Qs,
02.60.Lj}
\end{abstract}
\maketitle

Anharmonic oscillators continue to be of interest
because of their enormous phenomenological significance.$^{[1]}$
In particular, the quartic case has been studied extensively in different
contexts for classical and quantum mechanical systems, especially in
one- and two-dimensional spaces.$^{[2-7]}$  This is of particular interest
as it can help shed light on the structure of the corresponding
$\phi^{4}$ quantum field theory. Up until now, most of the results have
been based on a strong coupling expansion,$^{[8]}$ variationally modified
perturbation methods,$^{[9]}$ or lattice methods,$^{[10]}$. In each
case only numerical results are accessible.

Analytical solutions of the anharmonic oscillator problem for
polynomial potentials of degree $4k+2$, where $k$ is a positive
integer, and some model specific restrictions have been reported.
For example, the sextic and decatic cases have been studied rather
extensively.$^{[11-22]}$ The $k=1$ case was first studied in
[11,12]. This class of anharmonic oscillators is conditionally
solvable in a coordinate representation, in which finite polynomials multiplied
by an exponential can be adopted as the ansatz for wavefunctions.
The quasi-exactly solvability of these cases is explained in
detail in [23].  However, the important quartic case,
together with polynomials of degree $4k$ for $k\geq 1$
is absent from this list of quasi-exactly solvable cases on the full line.
The quartic anharmonic oscillator in one dimension on the half line
with potential $V(x)=\alpha x^{2}+\beta x^{4}$ and $x\geq 0$
was first studied by Flessas {\it et al} in [24, 25] in terms
of an integral expression.
It was commonly believed that the lowest-degree one-dimensional quasi-exactly
solvable polynomial potential is sextic.
This belief was based on the assumption that the Hamiltonian must be Hermitian. However,
it has been discovered recently that there are huge classes of non-Hermitian, parity and time reversal invariant (PT-symmetric) Hamiltonians whose spectra are real, discrete, and bounded below. Replacing hermiticity by the
weaker condition of PT symmetry allows for new kinds of quasi-exactly solvable theories.$^{[26-28]}$
In [26, 27], a new two-parameter family of quasi-exactly solvable
quartic polynomial potentials with
$V(x)=-x^{4}+2{\rm i}ax^{3}+(a^{2}- 2b)x^{2}+2{\rm i}(ab- J)x$,
where $a\neq 0$, $b\neq 0$, and $J$ is an integer,
is introduced.
In [28], the real spectrum of the potential $V(x)=x^2 (ix)^{\nu}$  with $\nu>0$
was studied.
In this paper, we show that exact solutions of the $O(N)$ invariant
quartic anharmonic oscillator in the
spherical coordinates can be exactly obtained when
the interaction parameter satisfies some special conditions
with an ansatz similar to that used for the sextic case.$^{[23]}$

\vskip .3cm

In $N$-dimensional space, the $O(N)$ invariant Hamiltonian for
the quartic anharmonic oscillator with unit mass can be written as

$$\hat{H}=-{1\over{2}}\sum_{i}{\partial\over{\partial x_{i}}}{\partial\over{\partial x_{i}}}+V({r}),\eqno(1)$$
where $V({r})$ is a polynomial in $r=\left(\sum_{i} x_{i}^{2}\right)^{1/2}$ of degree $4$.
Because of the underlying $O(N)$ symmetry, the Schr\"{o}dinger equation of (1)
can be transformed into the following form in $N$-dimensional spherical
coordinates:

$$\left(-{1\over{2}}\triangle+V(r)\right)\Psi({x_{i}})=E\Psi({x}_{i})\eqno(2)$$
with $\Psi({x}_{i})=R_{l}(r)H_{l}(\hat{\Omega})$ and

$$\triangle\Psi({x}_{i})=\left(
{d^{2}\over{dr^{2}}}+{N-1\over{r}}{d\over{dr}}-{l(l+N-2)\over{r^{2}}}\right)R_{l}(r)H_{l}(\hat{\Omega}).\eqno(3)$$
It will be shown that quasi-exact solutions to the potential
$$V(r)=\lambda_{1}r+\lambda_{2} r^{2}+\lambda_{4}r^{4}\eqno(4)$$
can be obtained in similar fashion to the polynomial potentials of degree $4k+2$
mentioned previously. It should be noted that the potential (4)
is directly related to the quantum double well
anharmonic oscillator in an external field studied in [29, 30].

 In order to solve the Schr\"{o}dinger equation (2)
with the potential (4), we use the ansatz:
$$R_{l}(r)=r^{l}\Phi(r)\exp\left(-\alpha r-\beta r^{3}/3\right),\eqno(5)$$
where $\alpha\neq0$ and $\beta>0$ are real parameters that are to be determined.
Similar to the polynomial potential of degree $4k+2$ cases, we
search for possible solutions to the potential (4), of which
$\Phi(r)$ is a finite polynomial.
In this case,
because the potential (4) is non-singular, the ansatz (5)
satisfies the standard physical requirement of normalizability of bound
states.$^{[31]}$  Namely, the ansatz (5) is finite at $r=0$ and always normalizable for $r\in[0,~\infty)$
even when $l=0$, of which the situation is similar to that in quasi-exactly solvable cases of
the polynomial potential of degree $4k+2$, but different from that for singular potentials
discussed in [31].

By substituting the ansatz (5) into (2), the differential equation for
$\Phi(r)$ is
$$-r{d^{2}\Phi(r)\over{d r^{2}}}+
(2\beta~r^3+2\alpha~ r-N-2 l+1 ){d\Phi(r)\over{d r}}
+$$$$\left[(2\lambda_1+( N+2 l+1)\beta )r^2-(2E+\alpha^2)r +(N+2 l-1)\alpha\right]\Phi(r)=0,\eqno(6)$$
in which the real parameters  $\alpha$ and $\beta$ should satisfy
$$\beta=\sqrt{2\lambda_4},~~\alpha=\lambda_{2}/\sqrt{2\lambda_4}\eqno(7)$$
in order to keep the polynomial quadratic in $r$ in front of $\Phi(r)$ in (6).
In search for polynomial solutions of Eq. (6), we write
$$\Phi(r)=\sum_{j}b_{j}r^{j},\eqno(8)$$
where $\{b_{j}\}$ are the expansion coefficients to be determined.
Substitution of (8) into Eq. (6) yields a matrix equation
for the expansion coefficients  $\{b_{j}\}$ and the corresponding
eigenenergies $E$. In general, the resulting matrix equation is
infinite dimensional. However, it can be shown that the matrix
truncates to one of finite order if the following
condition is satisfied:
$$\lambda_{1}=-(N+2l+ 2m+1)\beta/2~~~~{\rm for}~m=0,~1,~2,~\cdots. \eqno(9)$$
And in this case $\Phi(r)$ reduces to a polynomial of degree $m$.
It follows that the matrix equation for determining the expansion
coefficients $\{b_{j}\}$ in (8) is of the following
form:
$${\bf F}{\bf v}=0,\eqno(10)$$
where
$${\bf F}={\tiny\left(
\matrix{
(N+2l-1)\alpha&-(N+2l-1)&0\cdots\cr\cr
-2E-\alpha^{2}&(N+2l+1)\alpha~~&-2(N+2l)&0\cdots\cr\cr
-2m\beta&-2E-\alpha^{2}~~ ~&(N+2l+3)\alpha &-3(N+2l+1)&0\cdots
\cr\cr
0&-2(m-1)\beta&~~~~-2E-\alpha^{2}&(N+2l+5)\alpha &-4(N+2l+2)&0\cdots
\cr\cr
&\ddots~~~~~~~~~~~~~~~\ddots&~~~~~~~~~~~~~~~~\ddots&~~~~~~~~~~~~~~~~~~\ddots&~
~~~~~~~~~~~~~~~~~~~~~~~~~~\ddots
\cr\cr
&&-6\beta&~-2E-\alpha^{2}&~~(N+2l+2m-3)\alpha&-m(N+2l+m-2)\cr\cr
&&&-4\beta~&-2E-\alpha^{2}&(N+2l+2m-1)\alpha\cr\cr
&&&&-2\beta&-2E-\alpha^{2}\cr
}\right)}\eqno(11)$$\\
is a matrix with $m+2$ rows  and $m+1$ columns
and ${\bf v}=(b_{0},b_{1},\cdots,b_{m})^{\rm T}$, in which ${\bf b}^{\rm T}$
stands for transposition of ${\bf b}$.
With some further constraints on the parameters, solutions of (10)
provide quasi-exact solutions of (6).
The first row of ${\bf F}$ provides for two such conditions:
(1) $N=1-2l$ or (2) $N\neq 1-2l$ and $b_{1}=\alpha b_{0}$.
The first condition is only valid  for $N=1$ case with $l=0$,
while the second condition is valid for $N>1$ cases.
Therefore, when $N=1$, the following eigen-equation determines
possible solutions for Eq. (6):

$${\bf Pv}{=(2E+\alpha^2)}{\bf v},\eqno(12)$$
where

$${\bf P}={\tiny\left(
\matrix{
0&2\alpha~~&-2&0\cdots\cr\cr
-2m\beta&0&4\alpha &-6&0\cdots
\cr\cr
0&-2(m-1)\beta&~0&6\alpha &-12&0\cdots
\cr\cr
&\ddots~~~~~~~~~~~~~~~\ddots&~~~~~~~~~~~~~~~~\ddots&~~~~~~~~~~~~~~~~~~\ddots&~
~~~~~~~~~~~~~~~~~~~~~~~~~~\ddots
\cr\cr
&&-6\beta&~~~0&~~(2m-2)\alpha&-m(m-1)\cr\cr
&&&-4\beta~&0&2m\alpha\cr\cr
&&&&-2\beta&0\cr
}\right)}\eqno(13)$$\\
is a  $(m+1)\times (m+1)$ quadri-diagonal matrix.
Since the matrix ${\bf P}$ is non-symmetric,
its eigenvalues for $\beta>0$ and any real $\alpha$ may be complex and
therefore should be
discarded. The remaining solutions with real eigenvalue of ${\bf P}$
are possible physical solutions to the problem.
Therefore, the number of physical
solutions in this case may be less than $m+1$.
This case is similar to the quantum double well
anharmonic oscillator in an external field
discussed in [29, 30].

However, since parity is always conserved for the potential (4), the potential (4) should be
written explicitly as

$$V(x) =  \left\{ \begin{array}{cc} \lambda_{1}x+\lambda_{2}x^{2}+\lambda_{4}x^{4} \, , & ~~~~ x\geq0 \,   \\
-\lambda_{1}x+\lambda_{2}x^{2}+\lambda_{4}x^{4} \, , & ~~~~ x\leq 0 \,   \end{array} \right.  
\eqno(14)$$
when $N=1$. It is clear that the $N=1$ case provides
part of solutions for the potential (14) on the full line with $-\infty<x<+\infty$.\\

When $N>1$, the eigen-equation of ${\bf v}$ becomes

$${\bf Qv}=0,\eqno(15)$$
where

$${\bf Q}={\tiny\left(
\matrix{
\alpha&-1&0\cdots\cr\cr
-2E-\alpha^{2}&(N+2l+1)\alpha~~&-2(N+2l)&0\cdots\cr\cr
-2m\beta&-2E-\alpha^{2}~~ ~&(N+2l+3)\alpha &-3(N+2l+1)&0\cdots
\cr\cr
0&-2(m-1)\beta&~~~~-2E-\alpha^{2}&(N+2l+5)\alpha &-4(N+2l+2)&0\cdots
\cr\cr
&\ddots~~~~~~~~~~~~~~~\ddots&~~~~~~~~~~~~~~~~\ddots&~~~~~~~~~~~~~~~~~~\ddots&~
~~~~~~~~~~~~~~~~~~~~~~~~~~\ddots
\cr\cr
&&-6\beta&~-2E-\alpha^{2}&~~(N+2l+2m-3)\alpha&-m(N+2l+m-2)\cr\cr
&&&-4\beta~&-2E-\alpha^{2}&(N+2l+2m-1)\alpha\cr
}\right)},\eqno(16)$$\\
which is also a $(m+1)\times (m+1)$ quadri-diagonal matrix, together
with further constraint on parameters $\alpha$ and $\beta$:

$$\beta=-(E+{1\over{2}}\alpha^{2}){b_{m}\over{b_{m-1}}}.\eqno(17)$$
In this case, the condition $m\geq1$ must be satisfied.
Furthermore, because the parameter $\beta$ depends not only on $\alpha$,
but also on eigen-energy $E$ according to (17), similar to the $N=1$ case,
whether $\lambda_{2}<0$ or $\lambda_{2}>0$
should be satisfied in order to keep solutions of (15) and (17)
physical needs to be fixed on a case-by-case basis.
In the following, we list some eigen-energies for small $m$ and
the corresponding eigenvector ${\bf v}$ for $N=1$ and $N\geq2$
cases separately.

When $N=1$, there is only one eigenvalue with $E_{m=0}=-{1\over{2}}\alpha^{2}$, for which
both $\lambda_{2}>0$ and $\lambda_{2}<0$ are allowed;
there are two eigenvalues with

$$E_{m=1}^{(1)}=-{1\over{2}}\alpha^2 -\sqrt{-\lambda_{2}},~~~ b^{(1)}_1={\beta\over{\sqrt{-\lambda_{2}}}}b^{(1)}_{0},$$
 $$ E_{m=1}^{(2)}=-{1\over{2}}\alpha^2 +\sqrt{-\lambda_{2}},~~~b^{(1)}_1=- {\beta\over{\sqrt{-\lambda_{2}}}}b^{(1)}_0.
  \eqno(18)$$
Thus, $\lambda_{2}<0$ must be satisfied for $m=1$ as shown in (18).

When $N\geq2$, there is only one eigenvalue with
$E_{m=1}={1\over{2}}\alpha^{2}(N+2l)$ with $b_{1}=\alpha b_{0}$
and the constraint $\beta=-{1\over{2}}(N+2l+1)\alpha^3$,
for which the condition $\lambda_{2}<0$ must be satisfied;
there are two independent solutions for $m=2$ with either

$$E_{m=2}^{(1)} = -{1\over{8}}(N+2l+5+
    \sqrt{(N+2l+1)(9N+18l+25)})\alpha^2,$$$$
    b^{(1)}_2 ={5N+10l+5 +
    \sqrt{(N+2l+1)(9N+18l+25)}\over{
 8 (N+2 l)}}\alpha^2 b^{(1)}_{0},~~ b^{(1)}_1 = \alpha b^{(1)}_0,$$
 $$\beta={(N+2l+1+\sqrt{(N+2l+1)(9N+18l+25)})(5N+10l+5+ \sqrt{
 (N+2l+1)(9N+18l+25)})\alpha^3\over{64(N+2l)}},
 \eqno(19a)$$
 or

 $$E_{m=1}^{(2)} = -{1\over{8}}(N+2l+5-
    \sqrt{(N+2l+1)(9N+18l+25)})\alpha^2,$$$$
    b^{(2)}_2 ={5N+10l+5 -
    \sqrt{(N+2l+1)(9N+18l+25)}\over{
 8 (N+2 l)}}\alpha^2 b^{(2)}_{0},~~ b^{(2)}_1 = \alpha b^{(2)}_0,$$
 $$\beta={(N+2l+1-\sqrt{(N+2l+1)(9N+18l+25)})(5N+10l+5- \sqrt{
 (N+2l+1)(9N+18l+25)})\alpha^3\over{64(N+2l)}}.
 \eqno(19b)$$

Solving equations (12) or (15), we can also evaluate higher (lower) excited
states with conditions that the parameters should satisfy. As the
results are quite complicated, they will not be provided here.

In order to keep physical solutions only, one may project
the matrices ${\bf P}$ and ${\bf Q}$ onto the physical
subspace. Let ${\bf v}^{(i)}$ be $i$-th eigenvector
with the corresponding eigenvalue to be real, and
$\Lambda$ be the projection operator,
which can be expressed as

$$\Lambda= \sum_{i}{\bf v}^{(i)}({\bf v}^{(i)})^{\dagger}.\eqno(20)$$
Then, instead of (12) and (15), physical solutions for $N=1$ and $N>1$
cases can be obtained respectively from the following equations:

$$\Lambda{\bf P}\Lambda{\bf v}{=(2E+\alpha^2)}{\bf v},\eqno(21)$$

\noindent and
$$\Lambda{\bf Q}\Lambda{\bf v}=0.\eqno(22)$$

In addition, similar to the electrostatic interpretation of the location of
zeros of the Heine-Stieltjes polynomials$^{[32]}$ and the polynomials
for the sextic anharmonic oscillator,$^{[23]}$ the polynomial
$\Phi(r)$ satisfying Eq. (6) can also be determined in the same way.
Let the polynomial $\Phi_{m}(r)$, up to a normalization constant,
be expressed in terms of its zeros as

$$\Phi_{m}(r)=\prod^{m}_{i=0}(r-r_{i}).\eqno(23)$$

Since there is no node for $m=0$, and the first and second derivatives
of $\Phi_{0}(r)$ are zero,

$$\lambda_{1}=-{1\over{2}}(N+2l+1)\beta ,~~E_0=-{1\over{2}}\alpha^{2},~~
{\rm and}~~ N=1-2l\eqno(24)$$
must be satisfied according to Eq. (6), which is consistent with solution
for $N=1$ and $m=0$ case reported previously. Since the second derivative
of $\Phi_{1}(r)$ is zero, the condition

$$2\beta r_{1}^{2}+2\alpha-(N+2l-1){1\over{r_{1}}}=0\eqno(25)$$
must be satisfied. In this case, $\Phi_{1}(r)=r-r_{1}$.
It is obvious that there are two solutions of Eq. (25) with
$r_1=\pm\sqrt{-\alpha\over{\beta}}$ when $N=1-2l$.
Once can check that there is indeed a solution of Eq. (25)
with $r_1=-{1\over{\alpha}}$ when $N\neq 1-2l$ with condition
$\beta=-{1\over{2}}(N+2l+1)\alpha^3$, which must be satisfied in
this case for Eq. (6) with $r\neq r_1$. For $m\geq 2$,
one can write

$${\Phi^{\prime\prime}(r_{i})\over{ \Phi^{\prime}(r_{i})}}
=\sum_{j\neq i}{2\over{r_{i}-r_{j}}}\eqno(26)$$\\
at any zero $r_{i}$. Then, Eq. (6) at any zero $r_{i}$ becomes

$$\sum_{j\neq i}{2\over{r_{i}-r_{j}}}-2\beta r_{i}^{2}-2\alpha+{N+2l-1\over{r_{i}}}=0\eqno(27)$$
for $i=1,~2,~\cdots,~m$,
of which similar equations for the generalized Lame functions are known as the Niven equations.$^{[32]}$
Eq. (27) provides with important relations to determine the polynomial $\Phi_{m}(r)$.
Once the polynomial $\Phi_{m}(r)$ is determined, one can get the corresponding
energy and further constraints on the parameters from Eq. (6).
The results are similar to the solutions of the sextic case$^{[23]}$
and closely related to Bethe ansatz solutions of Lipkin-Meshkov-Glick model$^{[33,34]}$
and the nuclear pairing problem.$^{[35, 36]}$

\vskip .4cm
In summary, the $O(N)$ invariant quartic anharmonic oscillator
is shown to be exactly solvable if the interaction parameter
satisfies some special conditions.
The problem is directly related to that of a quantum double well
anharmonic oscillator in an external field. A finite dimensional
matrix equation for the problem is constructed explicitly, along with
analytical expressions for some excited states in the system.
The corresponding Niven equations for determining the polynomial
solutions for the problem are given. It is
expected that an ansatz similar to (5) may also be possible for finding
quasi-exactly solvable cases for anharmonic oscillators with
polynomials of degree $4k$ for $k>1$, which add another
quasi-exactly solvable class of anharmonic oscillators. These
results should be useful to test the validity of various
approximation method, to probe the true nature of the solution,
and to reveal its asymptotic behavior.

\vskip .4cm
Support from the U. S. National Science Foundation (PHY-0500291 \& OCI-0904874),
the Southeastern Universities Research Association, the Natural Science
Foundation of China (11175078), the Doctoral Program Foundation
of State Education Ministry of China (20102136110002),
and the LSU--LNNU joint research program (9961) is
acknowledged.

\end{document}